\documentclass[12pt]{article}
\usepackage{graphicx}

\begin{document}

\title{\textbf{A WIDE-FIELD CORRECTOR\\ AT THE PRIME FOCUS
OF A\\ RITCHEY--CHR\'ETIEN
TELESCOPE}\thanks{\textit{Astronomy~Letters,~Vol.~30,~No.~3,~2004,
~pp.~200--208.\qquad \qquad \qquad Translated from Pis'ma v
Astronomicheski$\breve \imath$ Zhurnal, Vol.~30, No.~3, 2004,
pp.~231--240. Original Russian text \copyright \,2004 by
Terebizh.}}}

\author{V.~Yu.~Terebizh\thanks{98409 Nauchny, Crimea, Ukraine;
 \,E-mail:\, \textsf{terebizh@crao.crimea.ua}}\\
 \small{\textit{Sternberg Astronomical Institute, Moscow, Russia}}}

\date{\footnotesize{Received July~16, 2003}}

\maketitle

\begin{quote}
\small{\textbf{Abstract}~--- We propose a form of a lens corrector
at the prime focus of a hyperboloidal mirror that provides a flat
field of view up to~$3^\circ$ diameter at image quality ${D_{80} <
0.8}$~arcsec in integrated (0.32--1.10~$\mu$m) light. The
corrector consists of five lenses made of fused silica. Only
spherical surfaces are used, so the system is capable of achieving
better images, if necessary, by aspherizing the surfaces. The
optical system of the corrector is stable in the sense that its
principal features are retained when optimized after significant
perturbations of its parameters. As an example, three versions of
the corrector are designed for the V.\,M.\,Blanco 4-m telescope at
Cerro Tololo Inter-American Observatory with $2^\circ.12$,
$2^\circ.4$, and $3^\circ.0$ fields of view.\\
 \copyright \, \textit{2004 MAIK ``Nauka/Interperiodica''.}

\medskip

}
\end{quote}

\newpage

\section*{Introduction}

The advent of reflectors with aperture diameters of 8~-- 10~m
required a revision of observational programs for telescopes of
preceding generations. Emphasis was placed on designing systems of
adaptive optics and realization sky surveys at the prime focus
with a wide-field corrector. The choice of the second way is
determined by several factors.

First of all, a few observational programs aimed at solving
important astrophysical problems, in particular, at studying
gamma-ray bursts, searching for hidden mass, and analyzing
gravitational lensing in clusters of galaxies, are of current
interest. For obvious reasons, the diameter of a Schmidt telescope
is difficult for increasing up to values well above the current
level of $\sim 1.3$~m. Special 4-m telescopes with lens correctors
designed together with the primary mirror are being proposed to
solve these problems: the Next Generation Lowell Telescope (NGLT)
(Blanco \emph{et al.}~2002) and the Visible and Infrared Survey
Telescope for Astronomy (VISTA) (McPherson \emph{et al.}~2002;
Emerson and Sutherland~2002). Particular attention is given to the
Large Synoptic Survey Telescope (LSST) with an effective aperture
of about 6.5~m at a primary mirror diameter of 8.4~m (Angel
\emph{et al.}~2000; Tyson~2002; Seppala~2002). At the same time,
being equipped with the prime-focus correctors with a field of
${\sim 1^\circ.5 - 2^\circ.0}$ diameter, the existing 4-m
Ritchey--Chr\'etien telescopes achieve an efficiency comparable to
the efficiency for the telescopes being designed.

Secondly, the fact that a field corrector to an existing telescope
can be made relatively fast also seems important.

Finally, at a diameter of $\sim 4$~m and a focal ratio of
$\sim2.5-3$, the primary mirror of a Ritchey--Chr\'etien telescope
with a roughly afocal field corrector can be matched in
modulation-transfer function with the main modern CCD detectors
with pixel sizes of $\sim 15$~$\mu$m. Thus, the challenging
problem during observations at the Cassegrain focus is solved in a
natural way.

% ======================================================= Fig.01
\begin{figure}[t]
   \centering
   \includegraphics[width=0.70\textwidth]{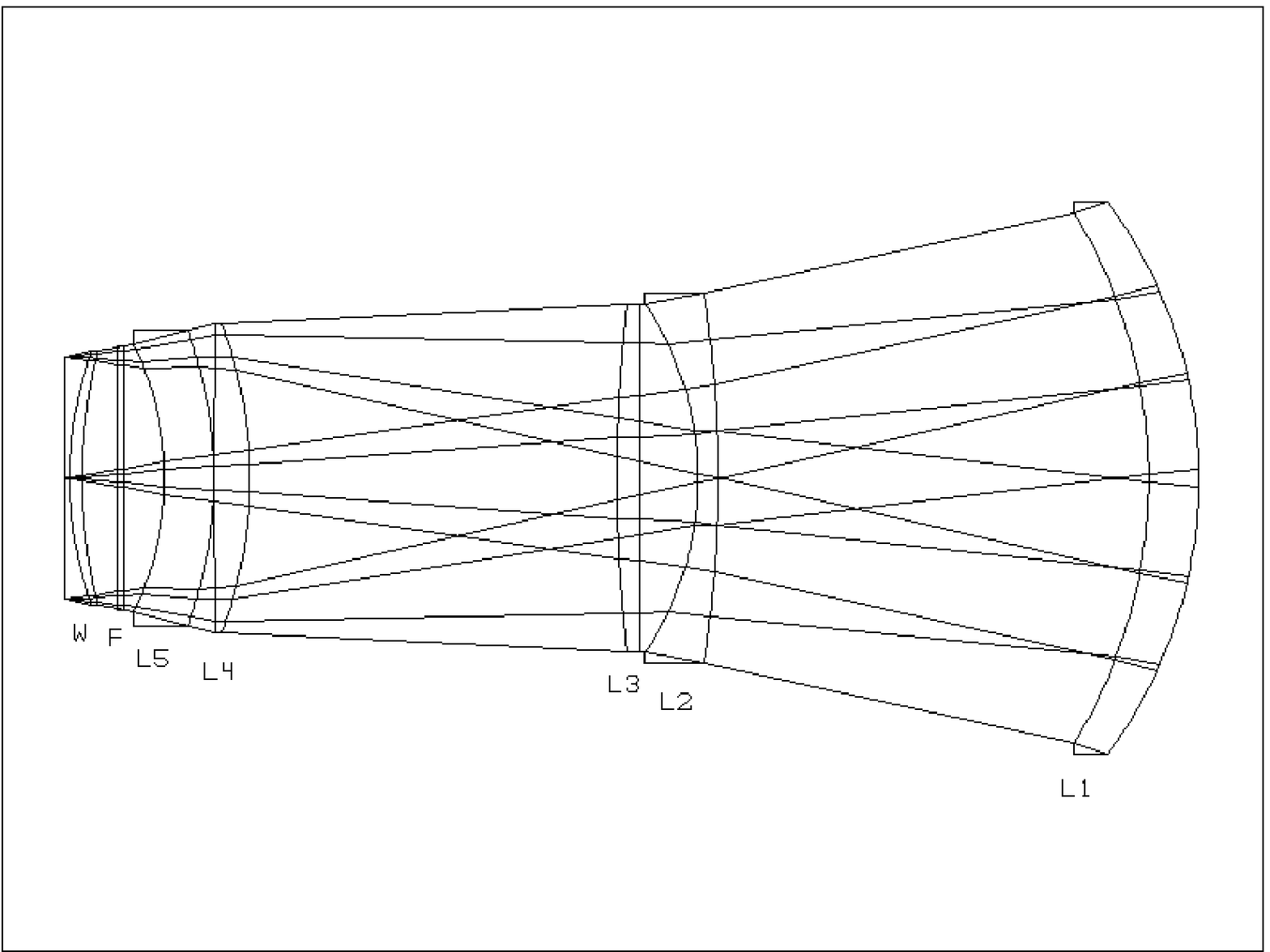}
   \caption*{Corrector ``S". The last elements are filter (F)
   and detector window (W).}
\end{figure}
% ================================================== End Fig.01

Much effort has been made to design field correctors at the prime
focus of a reflector (see the reviews by Wynne~1972;
Mikhelson~1976, \S7.5; Wilson~1996, \S4.3; Schroeder~2000, \S9.2).
The correctors designed by Ross~(1935) and Wynne~(1968) were the
systems that determined the development of this area of research
for a long time. The former corrector has a flat field of view of
${2w \simeq 15^{'}}$ diameter with stellar images better
than~$1^{''}$; the latter detector has an about~$50^{'}$ field of
similar quality. Many of the modern reflectors are equipped with
three-lens Wynne correctors or modifications of this system. The
observational programs being planned require a field of view no
less than $1^\circ.5$ in diameter. Thus, for example, the NGLT and
VISTA projects mentioned above are to provide a flat field of
$2^\circ$ diameter. The correctors designed for these telescopes
have four or five aspherical lenses, with the diameter of the
front lens reaching 1.25~m. The front lens of the LSST corrector
is of 1.34~m diameter; the concave surfaces of the lenses are
eighth-order aspherics.

Here, we propose a new type of corrector at the prime focus of a
hyperboloidal mirror (Fig.~1) that provides a flat field up
to~$3^\circ$ in diameter at image quality\footnote{The diameter of
a circle within which $80\%$ of energy in the image of a
point-like source is enclosed, $D_{80}$, is meant.} better than
$0^{''}.8$ in integrated (0.32--1.10~$\mu$m) light. Since
observations are generally carried out in relatively narrow
spectral ranges rather than in integrated light, the corresponding
image size with the corrector is smaller than this value. Only
spherical surfaces were used, so the system is not only relatively
easy to make, but also is capable of achieving better images by
aspherizing the surfaces. All lenses are made of fused silica.
This material has good manufacturing properties and provides high
transparency in the ultraviolet, which is often a key factor.

% ===================================================== Table 1
\begin{center}
 \footnotesize{
\begin{tabular}{|l|c|c|c|}
\multicolumn{4}{l}{\textbf{Table 1.} General characteristics of
the correctors}\\[5pt]
 \hline
 & \multicolumn{3}{|c|}{} \\
   \multicolumn{1}{|c|}{Parameter} &
   \multicolumn{3}{|c|} {Corrector} \\
   \cline{2-4}   \rule{0pt}{5mm}
 & $``R"$ & $``S"$ & $``T"$ \\[3pt]
 \hline     %\rule{0pt}{5mm}
 &&&\\
 Angular field of view, $2w$, deg & 2.12 & 2.4 & 3.0 \\[5pt]
 Effective focal length &&& \\
 with the telescope, mm  & 11 506.7 & 11 400.4 & 11 505.9 \\[5pt]
 Focal ratio & 2.92  & 2.90 & 2.92 \\[5pt]
 Scale, $\mu$m/arcsec & 55.79 & 55.27 & 55.78 \\[5pt]
 Linear field of view, mm & 427 & 481 & 606 \\[5pt]
 Spectral range, $\mu$m & 0.32--1.10 & 0.32--1.10 & 0.32--1.10 \\[5pt]
 Variation of image RMS-radius
    & 13.2--15.6 $\mu$m & 12.4--15.6 $\mu$m & 14.3--19.8 $\mu$m \\
 over field, 0.32--1.10 $\mu$m
    & $0^{''}.24-0^{''}.28$ & $0^{''}.22-0^{''}.28$ & $0^{''}.26-0^{''}.35$ \\[5pt]
 Variation of $D_{80}$ over field
    & 33.2--38.5 $\mu$m & 31.8--38.0 $\mu$m & 36.0--45.0 $\mu$m \\
 (center--edge, 0.32--1.10 $\mu$m)
    & $0^{''}.60-0^{''}.70$ & $0^{''}.58-0^{''}.68$ & $0^{''}.64-0^{''}.80$\\[5pt]
 Variation of $D_{80}$ over field
    & 20.0--39.3 $\mu$m & 17.4--40.2 $\mu$m & 19.6--52.8 $\mu$m \\
 in 0.35--0.45 $\mu$m band
    & $0^{''}.36-0^{''}.70$ & $0^{''}.32-0^{''}.72$ & $0^{''}.36-0^{''}.94$\\[5pt]
 Variation of $D_{80}$ over field
    & 24.4--30.3 $\mu$m & 20.2--25.8 $\mu$m & 24.0--28.2 $\mu$m\\
 in 0.54--0.66 $\mu$m band
    & $0^{''}.44-0^{''}.54$ & $0^{''}.37-0^{''}.47$ & $0^{''}.44-0^{''}.50$\\[5pt]
 Variation of $D_{80}$ over field
    & 25.8--38.3 $\mu$m & 20.4--33.8 $\mu$m & 25.4--38.8 $\mu$m \\
 in 0.70--0.90 $\mu$m band
    & $0^{''}.46-0^{''}.69$ & $0^{''}.37-0^{''}.61$ & $0^{''}.46-0^{''}.70$\\[5pt]
 Transmittance (including reflections, &&&\\
 without coatings, 0.32--1.10 $\mu$m)
 & 0.53--0.55 & 0.53--0.55 & 0.53--0.55 \\[5pt]
 Maximum distortion & 0.42\% & 0.60\% & 0.61\% \\[5pt]
 Maximum gradient &&&\\
 of distortion with wavelength, $\mu\mbox{m}^{-1}$
  & $3.88\times 10^{-4}$  & $2.25\times 10^{-4}$ & $3.25\times 10^{-4}$ \\[5pt]
 Types of lens surfaces & All spheres & All spheres & All spheres \\[5pt]
 Maximum clear aperture, mm & 900 & 1100 & 1300 \\[5pt]
\hline
\end{tabular}
 }
\end{center}
% ======================================================= End Table 1

\medskip

The proposed corrector system is designed for a hyperboloidal
mirror with a conic constant typical of Ritchey--Chr\'etien
telescopes. As specific examples, we discuss three versions of the
corrector for the V.\,M.\,Blanco 4-m telescope at Cerro Tololo
Inter-American Observatory~--- systems~$``R"$, $``S"$, and $``T"$,
with $2^\circ.12$, $2^\circ.4$, and $3^\circ.0$ fields of view,
respectively (Table~1). The last two systems should be considered
to be basic, while in system~$``R"$ designed for a reduced size of
the front lens, we had to introduce noticeable distortions.

\section*{Primary mirror of the telescope}

The parameters of the primary mirror of the Blanco telescope
(Table~2) were taken from the report by Gregory and Boccas~(2000).
The central obscuration is produced by a hole in the mirror and
stray-light baffles. Since images far from the diffraction limit
are dealt with in wide-field observations, the central obscuration
affects the images only slightly.

% ==================================================== Table 2
\begin{center}
 \footnotesize{
\begin{tabular}{|l|c|}
 \multicolumn{2}{l}{\textbf{Table 2.} Primary mirror}\\[5pt]
 \hline
 \multicolumn{1}{|c|}{Parameter} & Value\\
 \hline
 &\\
 Radius of curvature at vertex & $-21\,311.6$ mm \\[5pt]
 Conic constant $k$            & $-1.09763$ \\[5pt]
 Aperture diameter             & $3934$ mm \\[5pt]
 Central obscuration           & $1651$ mm \\[5pt]
\hline
\end{tabular}
 }
\end{center}
% ================================================= End Table 2

Note that the correctors described below need to be adjusted only
slightly for a moderate variation of the parameters of the primary
mirror given in Table~2. In particular, this is true for a
paraboloidal primary mirror.

\section*{System $``\mbox{\textbf{S}}"$}

The layout of corrector~$``S"$ is shown in Fig.~1 (for detailed
information, see Table~3). The letters~FS denote fused silica. The
most commonly used Schott~BK7 glass was taken as the material for
the filter; clearly, the system is not critical in this regard,
and choosing a different glass as well as adopting a different
thickness of the filter can be easily compensated.

If we remove lenses~L3 and~L5 from system~$``S"$, then the
remaining part will resemble the classical system by Wynne~(1968).
Necessity of addition of these two lenses to produce a really
large field of view is caused by the fact that, in this case,
doublets~L2$+$L3 and~L4$+$L5 are formed; each of them effectively
suppresses the aberrations of the primary mirror, first of all,
coma. It is interesting to note, in this connection, that the
lens~L3 has already appeared in the corrector designed by
Delabre~(2002) for a ${2w = 0^\circ.95}$ field of view. Delabre's
system consists of three lenses and a detector window which has an
optical power.

% ============================================= Table 03
\begin{center}
 \footnotesize{
\begin{tabular}{|c|l|c|c|c|c|}
 \multicolumn{6}{l}{\textbf{Table 3.} Design data for the
 system $``S"$}\\[5pt]
\hline
  & & Radius & & & Clear \\
 Surface & \multicolumn{1}{|c|}{Comments} & of curvature, &
  Thickness, & Glass & aperture, \\
 number & & mm & mm & & mm \\
\hline
 &&&&&\\
 1  & Aperture stop  &$\infty$ &90.755 &---   &3934.00 \\
 2  &Primary mirror  &&&&\\
    &($k=-1.09763$)  &$-21311.6$ &$-8521.90$ &Mirror &3934.00\\
 3  &L1       &$-921.47$   &$-100.00$ &FS     &1100.00 \\
 4  &         &$-1017.83$  &$-855.28$ &---    &1056.78 \\
 5  &L2       &$-2321.62$  &$-40.00$  &FS     &740.03 \\
 6  &         &$-620.63$   &$-116.04$ &---    &693.91 \\
 7  &L3       &$\infty$    &$-45.00$  &FS     &693.96 \\
 8  &         &$3077.69$   &$-730.57$ &---    &694.04 \\
 9  &L4       &$-872.53$   &$-70.13$  &FS     &619.93 \\
 10 &         &$33728.22$  &$-1.00$   &---    &616.39 \\
 11 &L5       &$-865.49$   &$-98.20$  &FS     &591.30 \\
 12 &         &$-620.76$   &$-80.19$  &---    &533.31 \\
 13 &Filter   &$\infty$    &$-12.00$  &BK7    &526.08 \\
 14 &         &$\infty$    &$-71.36$  &---    &523.30 \\
 15 &Window   &$1047.66$   &$-25.00$  &FS     &508.85 \\
 16 &         &$734.15$    &$-10.00$  &---    &508.06 \\
 17 &Detector &$\infty$    &          &       &480.55 \\
\hline
\end{tabular}
 }
\end{center}
% ============================================= End Table 03

\medskip

An optical power is also planned to be imparted to the detector
window in the correctors described here. A slightly worse, but
comparable image quality is achieved for a flat window. However,
it seems natural to use additional degrees of freedom, given the
total number of optical surfaces.

The five-lens system shown in Fig.~1 is \emph{stable} in the sense
that its principal features are retained when optimized after
significant perturbations of its parameters. The final state in
stable systems is reached abruptly; i.e., either a global or a
nearly global minimum of the merit function is realized in the
multi-dimensional space of optical parameters. The numerous
variations of a three-lens corrector show that a similar stability
is also characteristic of Wynne's triplet, but in lower-dimension
space. These features of the five-lens system allow it to be
considered as a new type of field corrector at the prime focus of
a reflector.

%------------------------------------------------ Fig.02
\begin{figure}[t]
   \centering
   \includegraphics[width=0.80\textwidth]{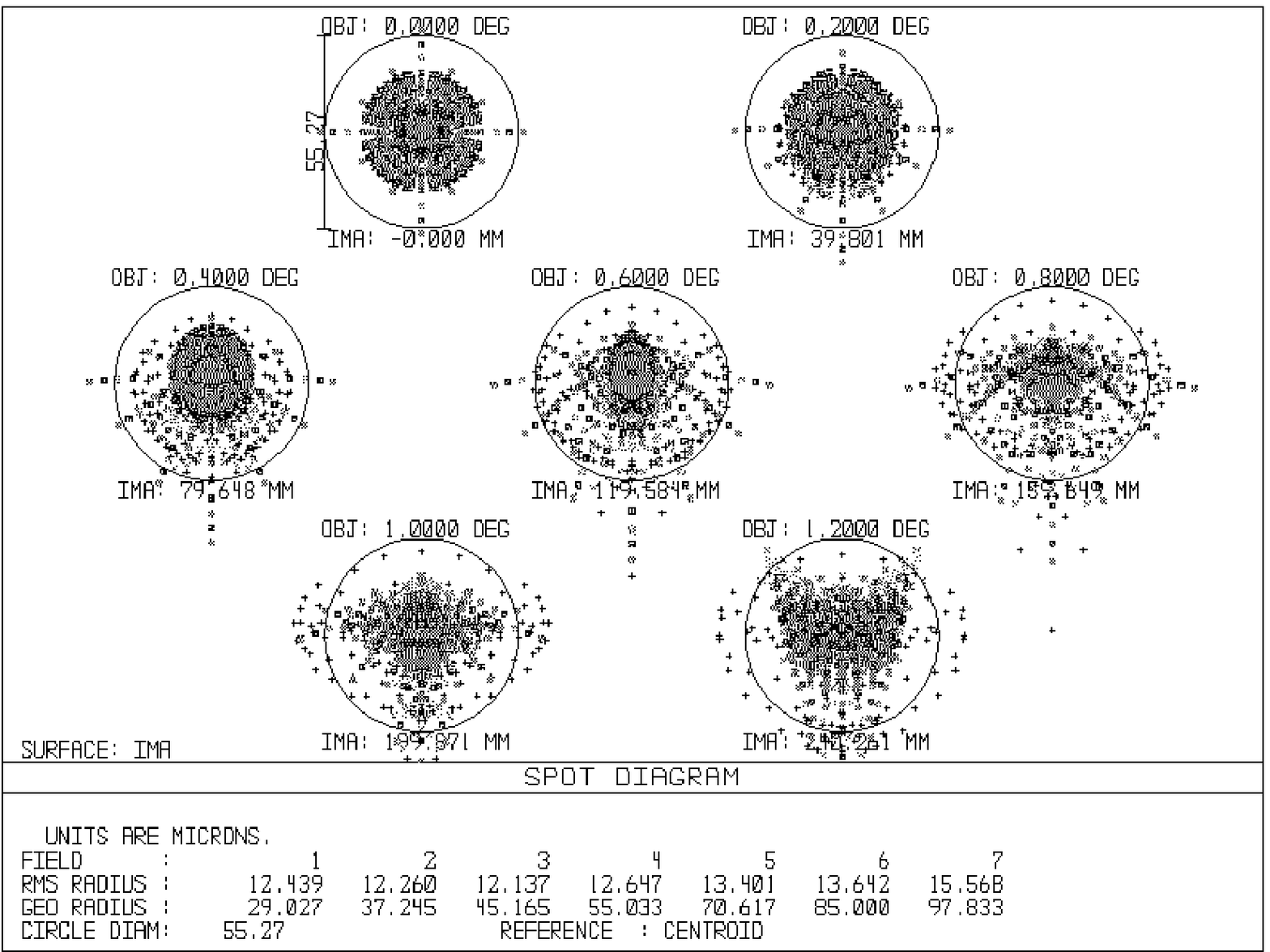}
   \caption*{Spot diagram for corrector ``S" over the range
   0.32--1.10~$\mu$m for field angles of $0$, $0^\circ.2$,
   $0^\circ.4$, $0^\circ.6$, $0^\circ.8$, $1^\circ.0$, and
   $1^\circ.2$. The circle diameters correspond to 1~arcsec
   (55.27~$\mu$m). The root-mean-square (RMS) and geometrical (GEO)
   radii of the images of a point-like source (in $\mu$m) are
   indicated for each of seven field angles.}
\end{figure}
%-------------------------------------------- End Fig.02
%------------------------------------------------ Fig.03
\begin{figure}[t]
   \centering
   \includegraphics[width=0.90\textwidth]{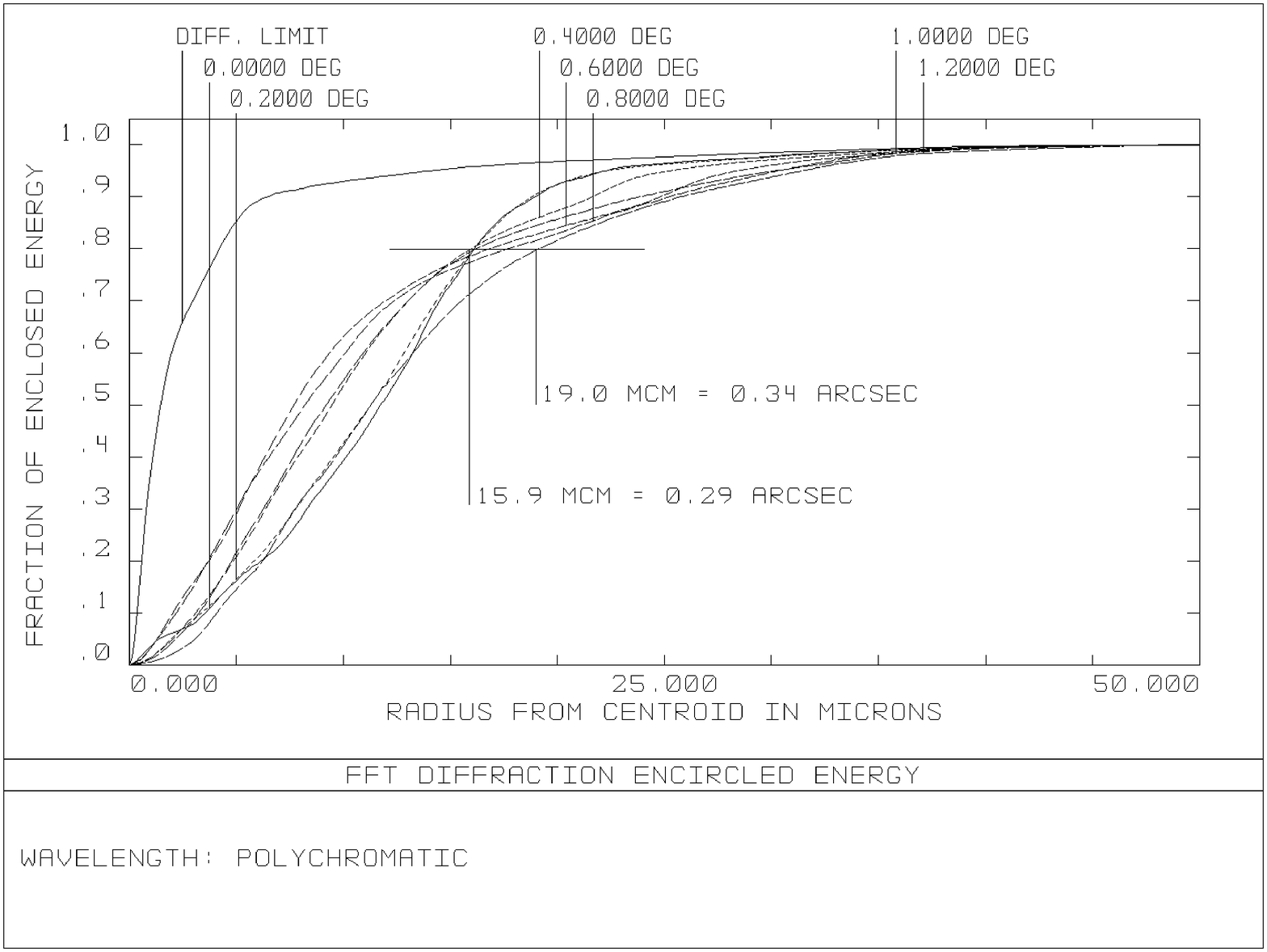}
   \caption*{Integral energy distribution along the radius in the
   diffraction stellar image for corrector ``S" in the
   range 0.32--1.10~$\mu$m for field angles of $0$, $0^\circ.2$,
   $0^\circ.4$, $0^\circ.6$, $0^\circ.8$, $1^\circ.0$, and
   $1^\circ.2$. The $80\%$ level and the corresponding extreme
   values of the radius (in $\mu$m and arcseconds) are indicated.}
\end{figure}
%-------------------------------------------- End Fig.03
%------------------------------------------------ Fig.04
\begin{figure}[t]
   \centering
   \includegraphics[width=0.70\textwidth]{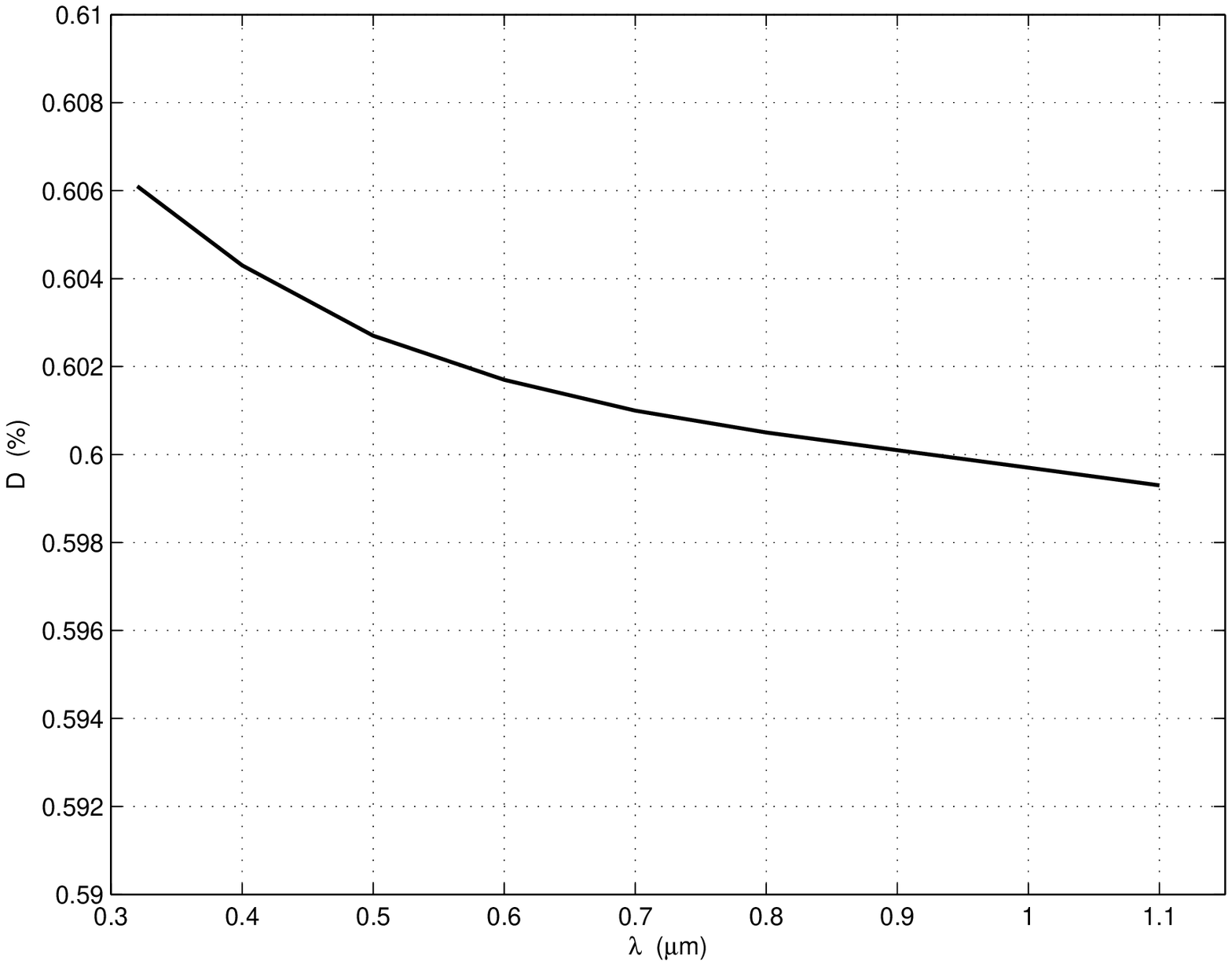}
   \caption*{Relative image distortion versus wavelength at the
   edge of the field of system ``S".}
\end{figure}
%-------------------------------------------- End Fig.04
%------------------------------------------------ Fig.05
\begin{figure}[t]
   \centering
   \includegraphics[width=0.80\textwidth]{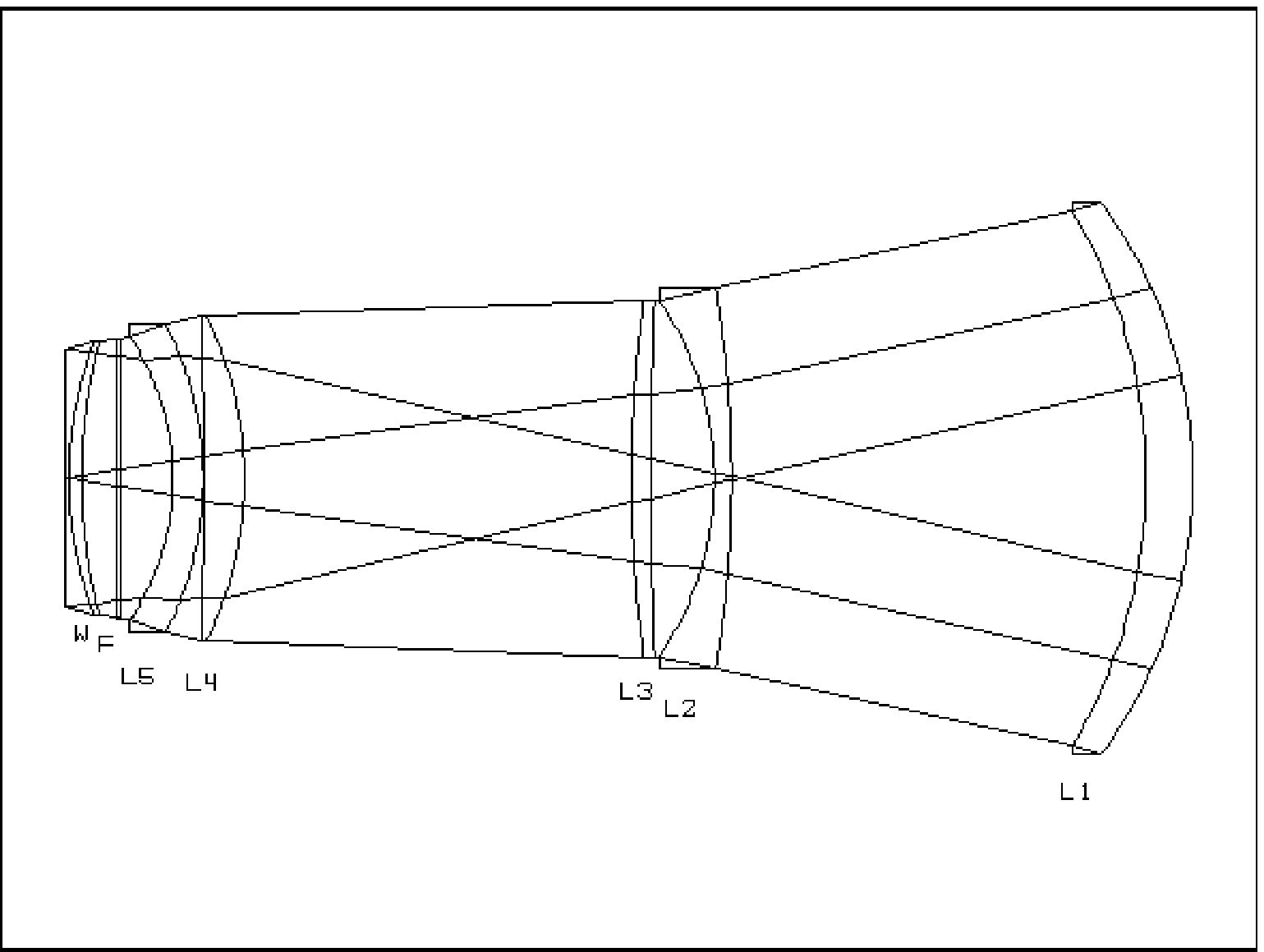}
   \caption*{Corrector ``T". The last elements are filter (F)
   and detector window (W).}
\end{figure}
%-------------------------------------------- End Fig.05
%------------------------------------------------ Fig.06
\begin{figure}[t]
   \centering
   \includegraphics[width=0.90\textwidth]{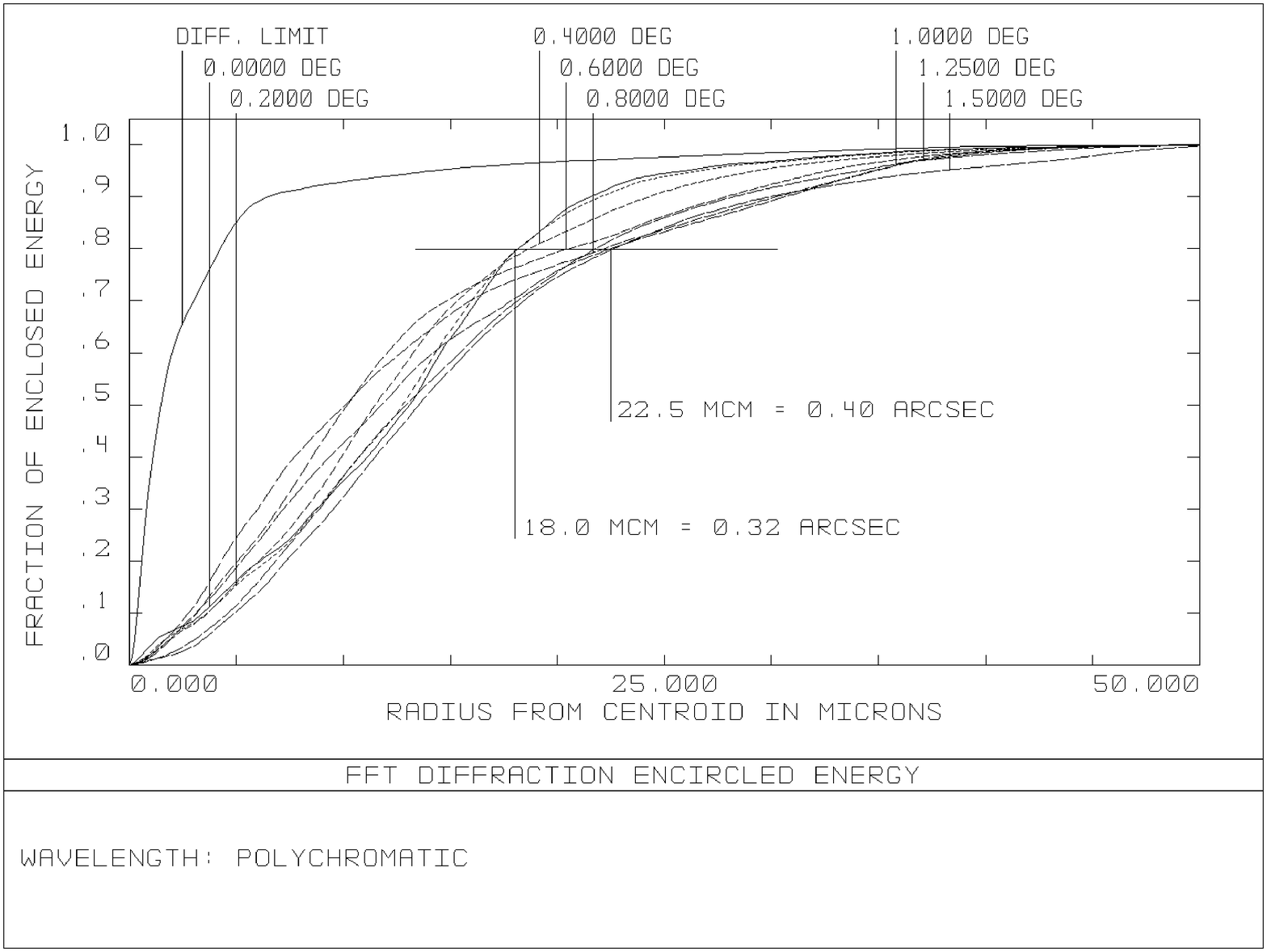}
   \caption*{Integral energy distribution along the radius in the
   diffraction stellar image for corrector ``T" in the
   range 0.32--1.10~$\mu$m for field angles of $0$, $0^\circ.2$,
   $0^\circ.4$, $0^\circ.6$, $0^\circ.8$, $1^\circ.0$, $1^\circ.25$,
   and $1^\circ.5$. The $80\%$ level and the corresponding extreme
   values of the radius (in $\mu$m and arcseconds) are indicated.}
\end{figure}
%-------------------------------------------- End Fig.06

Figure~2 shows the spot diagram\footnote{The distribution of light
rays in a stellar image on the focal plane.} for system~$``S"$ in
integrated light. A slightly clearer idea of the image quality can
be got from the plot of the fraction of enclosed energy in the
diffraction stellar image shown in Fig.~3. We took special
measures in order that the diameters of the stellar images remain
constant over the entire field. As one can see from Fig.~3,
$D_{80}$ changes from~$0^{''}.58$ on the optical axis
to~$0^{''}.68$ at the edge of the field. The image quality in
narrow spectral bands is given in Table~1. Note that the
refocusing range when passing from one spectral band to another is
only a few hundredths of a millimeter; such a small value is
attributable to the optimization of the system in integrated
light.

The corrector is close to an afocal system, so the focal length of
the telescope exceeds the focal length of the primary mirror only
slightly (see Tables~1 and~2). At a pixel size of $\sim 15$~$\mu$m
characteristic of the CCD detectors currently used in astronomy,
an angle of~$0^{''}.27$ corresponds to one pixel in the focal
plane. Thus, approximately 1.5 to 2.5 pixels fit into the
diameter~$D_{80}$, depending on the width of the spectral range
used in observations. This matching of the optical system with the
detector should be considered to be satisfactory.

Since the lenses of the corrector are made of fused silica, its
transmittance depends weakly on wavelength in the spectral range
0.32--1.10~$\mu$m considered here (the data of Table~1 refer to
lenses L1--L5). The deposition of effective modern coatings will
provide a ${\sim 83\%}$ transmittance of the corrector. Thus,
addition of two fused silica lenses to Wynne's corrector, which
will affect the system transparency only slightly, allows us to
significantly expand the field of view with good images.

For the linear fields of view of the order half a meter of
interest, not so much the image distortion as its variation with
wavelength is hazardous (Ingerson~1997). In system~$``S"$, the
positive\footnote{Often called \textit{pincushion}.} distortion
reaches its maximum at the edge of the field of view in the
ultraviolet; the exact value is $0.606\%$ for a field angle of ${w
= 1^\circ.2}$ and a wavelength of 0.32~$\mu$m. If we do not pose a
special astrometric problem, then this value may be considered
negligible. Otherwise, being constant with time, the distortion
can be taken into account when processing the data.

In our case, the variation of the distortion with wavelength is
shown in Fig.~4. The maximum (in absolute value) distortion
gradient is $-2.25\times 10^{-4}$~$\mu\mbox{m}^{-1}$. Suppose, for
example, that the observations are carried out in a
0.1-$\mu$m-wide ultraviolet band. The length of the spectrum
attributable to the distortion variation with wavelength is then
5.4~$\mu$m (the radius of the field of view was taken to be
240~mm). This length is only a small fraction of the total image
size $D_{80} \simeq 40$~$\mu$m at the edge of the field in the
ultraviolet. Since the effect under discussion plays an
appreciably lesser role in other spectral ranges, the distortion
in system~$``S"$ may be considered acceptable.

\section*{System $``\mbox{\textbf{T}}"$}

The layout of corrector~$``T"$ is shown in Fig.~5; its parameters
are given in Table.~4.

In this case, the image quality (Fig.~6) is only slightly worse
than that for system~$``S"$. As above, an angle of $0^{''}.27$
corresponds to a 15-$\mu$m detector pixel in the focal plane.
Therefore, the above remarks concerning the matching of the
optical system and the light detector remain valid.

As far as the basic optical system remained virtually unchanged
when the field of view increased significantly, from~$2^\circ.4$
up to~$3^\circ.0$, it primarily corresponds just to wide-field
observations. In fact, the introduction of two doublets
effectively suppressing coma pursued this goal. It is still
possible to find a four-lens corrector with an image quality that
is only slightly worse than that for the five-lens system~$``S"$,
but we failed to find a four-lens ``double'' of system~$``T"$ with
its $3^\circ$ field.

The refocusing range of the corrector when changing the spectral
band does not exceed 0.05~mm.

The transparency of the system under consideration in the entire
spectral range 0.32--1.10~$\mu$m is virtually the same as that for
system~$``S"$.

The type of image distortion (positive) was also preserved. At the
edge of the field of view, the distortion slightly increases from
$0.607\%$ in the ultraviolet to $0.611\%$ in the infrared. The
distortion in both correctors is small; if necessary, an
orthoscopic corrector can be designed rigorously. The largest
distortion gradient with wavelength, namely, $3.25\times
10^{-4}$~$\mu\mbox{m}^{-1}$, is reached at the edge of the field
again for ${\lambda = 0.32}$~$\mu$m. At observations in a
0.1-$\mu$m-wide ultraviolet band, the distortion variation with
wavelength causes the images to blur at the edge of the field by
slightly less than 10~$\mu$m. This value is almost twice as large
as the value for system~$``S"$. However, as above, it is small
compared to the sizes of the images themselves.

% =================================================== Table 04
\begin{center}
 \footnotesize{
\begin{tabular}{|c|l|c|c|c|c|}
 \multicolumn{6}{l}{\textbf{Table 4.} Design data for the
 system $``T"$}\\[5pt]
\hline
  & & Radius & & & Clear \\
 Surface & \multicolumn{1}{|c|}{Comments} & of curvature, &
  Thickness, & Glass & aperture, \\
 number & & mm & mm & & mm \\
\hline
 &&&&&\\
 1  & Aperture stop  &$\infty$ &90.755 &---   &3934.00 \\
 2  &Primary mirror  &&&&\\
    &($k=-1.09763$)  &$-21311.6$ &$-8150.16$  &Mirror &3934.00\\
 3  &L1       &$-1084.73$  &$-110.00$ &FS     &1300.00 \\
 4  &         &$-1226.17$  &$-975.325$&---    &1256.42 \\
 5  &L2       &$-2754.95$  &$-40.00$  &FS     &895.21 \\
 6  &         &$-742.97$   &$-150.24$ &---    &841.71 \\
 7  &L3       &$9276.47$   &$-45.00$  &FS     &841.83 \\
 8  &         &$3011.10$   &$-910.86$ &---    &842.74 \\
 9  &L4       &$-866.96$   &$-96.00$  &FS     &772.69 \\
 10 &         &$-14737.23$ &$-1.00$   &---    &769.07 \\
 11 &L5       &$-792.85$   &$-71.00$  &FS     &726.26 \\
 12 &         &$-602.28$   &$-121.60$ &---    &669.50 \\
 13 &Filter   &$\infty$    &$-12.00$  &BK7    &661.82 \\
 14 &         &$\infty$    &$-81.44$  &---    &658.87 \\
 15 &Window   &$1269.90$   &$-30.00$  &FS     &643.51 \\
 16 &         &$901.66$    &$-10.00$  &---    &642.55 \\
 17 &Detector &$\infty$    &          &       &606.38 \\
\hline
\end{tabular}
 }
\end{center}
% ================================================ End Table 04

\medskip

\section*{System $``\mbox{\textbf{R}}"$}

In the Introduction, we noted that decreasing the diameter of the
front lens in system~$``R"$ entails a noticeable distortion of the
basis system. For this reason, system~$``R"$ is given here as a
supplement to the correctors considered above. Nevertheless,
system~$``R"$, taken in itself, is of interest in realizing a
field of view slightly larger than~$2^\circ$ (the adopted specific
field diameter of~$2^\circ.12$ corresponds to the diagonal of a
square with a $1^\circ.5$ side).

Table~1 gives a description of system~$``R"$ enough to get an idea
of the image quality. Table~5 lists parameters of the optical
elements; further information can be obtained after inputting the
data of Table~5 to some optical program.

As we see from Table~1, the relatively compact system~$``R"$
provides roughly the same image quality as does system~$``S"$, but
within a somewhat smaller field of view.

% =============================================== Table 05
\begin{center}
 \footnotesize{
\begin{tabular}{|c|l|c|c|c|c|}
 \multicolumn{6}{l}{\textbf{Table 5.} Design data for the
 system $``R"$}\\[5pt]
\hline
  & & Radius & & & Clear \\
 Surface & \multicolumn{1}{|c|}{Comments} & of curvature, &
  Thickness, & Glass & aperture, \\
 number & & mm & mm & & mm \\
\hline
 &&&&&\\
 1  & Aperture stop  &$\infty$ &90.755 &---   &3934.00 \\
 2  &Primary mirror  &&&&\\
    &($k=-1.09763$)  &$-21311.6$ &$-9011.089$ &Mirror &3934.00\\
 3  &L1       &$-705.30$   &$-81.34$  &FS     &900.00 \\
 4  &         &$-823.90$   &$-497.65$ &---    &869.64 \\
 5  &L2       &$-1483.06$  &$-35.16$  &FS     &670.60 \\
 6  &         &$-506.93$   &$-304.37$ &---    &621.38 \\
 7  &L3       &$14674.93$  &$-42.38$  &FS     &614.17 \\
 8  &         &$1829.42$   &$-472.97$ &---    &613.92 \\
 9  &L4       &$-385.00$   &$-35.00$  &FS     &512.86 \\
 10 &         &$-382.33$   &$-51.69$  &---    &494.85 \\
 11 &L5       &$-745.23$   &$-62.00$  &FS     &494.24 \\
 12 &         &$-1168.35$  &$-44.72$  &---    &478.18 \\
 13 &Filter   &$\infty$    &$-12.00$  &BK7    &471.52 \\
 14 &         &$\infty$    &$-70.04$  &---    &468.96 \\
 15 &Window   &$878.97$    &$-25.22$  &FS     &455.64 \\
 16 &         &$596.19$    &$-10.00$  &---    &455.21 \\
 17 &Detector &$\infty$    &          &       &427.71 \\
\hline
\end{tabular}
 }
\end{center}
% ============================================ End Table 05

\medskip

\section*{Concluding remarks}

The field correctors considered here are relatively simple
systems: the lens surfaces are spherical, and the lenses
themselves are made of the same material. The question as to
whether to aspherize some or all of the surfaces should probably
be solved depending on specific circumstances that include the
need for achieving better images, the corrector production cost,
etc.

Just as it occurs in adaptive optics systems, the requirement of
providing high transmittance of the system in the outermost
ultraviolet range 0.32~-- 0.34~$\mu$m presents the greatest
difficulty in using the lenses (see, in particular, Tokovinin
\emph{et al.}~2003). Since the total thickness of the corrector
lenses is large, this requirement, in fact, narrows down the
choice to one material~--- fused silica. Quartz optics is known to
be transparent far beyond the range 0.32--1.10~$\mu$m considered
here. From the optical point of view, it becomes possible to use a
single material in such a complex system as the field corrector,
because a moderate change in the focal length of the telescope is
admissible.

Apart from high transparency, there are also other reasons for
seeking to make the system purely of fused silica:
\begin{itemize}
\item[\textbf{--}] reliable manufacturing procedures for producing
large homogeneous blanks of this material have now been developed;
\item[\textbf{--}] fused silica is well polished;
\item[\textbf{--}] it firmly holds coatings;
\item[\textbf{--}] all lenses of the system have not only a small,
but also the same thermal expansion coefficient\footnote{According
to Schott Lithotec~(2003), it is $0.5\times 10^{-6}/\mbox{K}$ in
the range $25^\circ - 100^\circ\mbox{C}$.};
\item[\textbf{--}] fused silica has a good time stability.
\end{itemize}
A discussion of attendant questions can be found in~\S3.3 of the
monograph by Wilson~(1999).

Observations with telescopes that provide subarcsecond images need
to be corrected for differential atmospheric refraction (Wynne and
Worswick~1986; Wynne~1986; Wilson~1996, \S4.4; Schroeder~2000,
\S9.5). The corresponding atmospheric dispersion corrector (ADC)
can be realized by making the lenses of the field corrector more
complex. In recent years, however, an ADC has been customarily
built into the field corrector as an independent device.

At first glance, systems of the type described here have a linear
field of view that is too large for CCD detectors to be
effectively used. Thus, for example, the diameter of the field of
view is 481~mm for system~$``S"$ and exceeds 600~mm for
system~$``T"$. Meanwhile, such field sizes are typical in the
modern projects of wide-field telescopes. For example, a linear
field of 550~mm diameter is expected to be achieved in the LSST
project. The main difficulty here is associated not with the
covering of the focal plane with many CCD chips, but with the
necessity of rapidly reading out and promptly processing an
extremely large amount of information. This problem has already
been solved in some of the existing systems (Lesser and
Tyson~2002; Walker~2002).

\section*{Acknowledgments}

I am grateful to V.\,V.\,Biryukov (Moscow State University) and
A.\,A.\,Tokovinin (Cerro Tololo Inter-American Observatory) for
valuable discussions of questions related to the use of large
telescopes.

\vspace{0.2cm} \hspace{5cm}
 \textit{Translated by V.~Astakhov}


\begin{thebibliography}{}

   \bibitem{} J.R.P.Angel, M.Lesser, R.Sarlot, and
    T.Dunham, ASP Conf. Ser.~\textbf{195}, 81 (2000).

   \bibitem{} D.R.Blanco, G.Pentland, C.H.Smith,
    T.Dunham, and R.L.Millis, Proc. SPIE No.~4842-20 (2002).

   \bibitem{} B.Delabre, \textit{Optical Design for Astronomical
    Instruments} (Rio de Janeiro),
    http://www.on.br/institucional/portuguese/ciclo2002/pub/ Delabre/
    RIO2002b.PPT (2002).

   \bibitem{} J.P.Emerson and W.Sutherland, Proc. SPIE No.~4836-08
    (2002).

   \bibitem{} B.Gregory and M.Boccas, \textit{The Blanco 4-m Telescope},
    http:// www.ctio.noao.edu/telescopes/4m/4m.html (2000).

   \bibitem{} T.E.Ingerson, \textit{Empirical and Theoretical
    Modeling of the PFADC Corrector on the Blanco 4-m Telescope},
    http://www.ctio.noao.edu/ telescopes/4m/pfadc/pfadc\_tei.html
    (1997).

   \bibitem{} M.P.Lesser and J.A.Tyson, Proc. SPIE No.~4836-38 (2002).

   \bibitem{} Schott Lithotec, \textit{Synthetic Fused Silica.
    Optical and Technical Grades} (2003).

   \bibitem{} A.McPherson, S.C.Craig, and W.Sutherland, Proc. SPIE
    No.~4837-10 (2002).

   \bibitem{} N.N.Mikhelson, \textit{Optical Telescopes. Theory
    and Design}, Moscow, Nauka, (1976) [in Russian].

   \bibitem{} F.E.Ross, Astrophys. J. \textbf{81}, 156 (1935).

   \bibitem{} L.G.Seppala, Proc. SPIE No.~4836-19 (2002).

   \bibitem{} D.J.Schroeder, \textit{Astronomical Optics},
    Academic, San Diego (2000).

   \bibitem{} A.Tokovinin, B.Gregory, H.E.Schwarz, V.Terebizh,
    S.Thomas, Proc. SPIE \textbf{439}, 673 (2003).

   \bibitem{} J.A.Tyson, Proc. SPIE No.~4836-04 (2002).

   \bibitem{} A.R.Walker, Mem. Soc. Astron. Ital. \textbf{73}, 23
    (2002).

   \bibitem{} R.N.Wilson, \textit{Reflecting Telescope Optics},
    Springer, Berlin (1996), Vol.~I.

   \bibitem{} R.N.Wilson, \textit{Reflecting Telescope Optics},
    Springer, Berlin (1999), Vol.~II.

   \bibitem{} C.G.Wynne, Astrophys. J. \textbf{152}, 675 (1968).

   \bibitem{} C.G.Wynne, \textit{Progress in Optics}, Ed. by
    E.Wolf, North-Holland, Amsterdam (1972), Vol.~10, p.~139.

   \bibitem{} C.G.Wynne, The Observatory \textbf{106}, 163
    (1986).

   \bibitem{} C.G.Wynne and S.P.Worswick, Mon. Not. R. Astron.
    Soc. \textbf{220}, 657 (1986).

\end{thebibliography}
\end{document}